\documentclass[preprint,a4paper]{article}

\usepackage{amssymb}

\newtheorem{theorem}{Theorem}

\newtheorem{proposition}{Proposition}
\newtheorem{corollary}{Corollary}

\newtheorem{example}{Example}
\newtheorem{definition}{Definition}

\begin{document}

\title{Canonical binary matrices related to bipartite graphs}

\author{Krasimir Yordzhev}
\date{\empty}
\maketitle
\begin{center}
Faculty of Mathematics and Natural Sciences\\
South-West University ''Neofit Rilski''\\
2700 Blagoevgrad, Bulgaria\\
E-mail: {yordzhev@swu.bg}
\end{center}

\begin{abstract}
The current paper is dedicated to the problem of finding the number of mutually non isomorphic bipartite graphs of the type  $g=\langle R_g ,C_g ,E_g \rangle$ at given $n=|R_g |$ and $m=|C_g |$, where $R_g$ and $C_g$ are the two disjoint parts of the vertices of the graphs $g$, and $E_g$ is the set of edges, $Eg \subseteq R_g \times C_g$. For this purpose,  the concept of canonical binary matrix is introduced. The different canonical matrices unambiguously describe the different with exactness up to isomorphism bipartite graphs. We have found a necessary and sufficient condition an arbitrary matrix to be canonical. This condition could be the base for realizing recursive algorithm for finding all $n \times m$ canonical binary matrices and consequently for finding all with exactness up to isomorphism binary matrices with cardinality of each part equal to $n$ and $m$.
\end{abstract}



\section{Introduction  and notation}

Let $k$ and $n$ be positive integers, $k\le n$. By $[n]$ we denote the set $[n] =\left\{ 1,2,\ldots ,n\right\}$ and by $[k,n]$ the set $[k,n] =\left\{ k,k+1,\ldots ,n\right\}$.

\emph{Bipartite graph} is the ordered triplet
 $$
g=\langle R_g , C_g , E_g \rangle ,
$$
where $R_g$ and $C_g$ are non-empty sets such that $R_g \cap C_g =\emptyset$, the elements of which will be called \emph{vertices}. $E_g \subseteq R_g \times C_g =\{ \langle r,c \rangle \; |\; r\in R_g ,c\in C_g \}$  - the set of \emph{edges}. Multiple edges are not allowed in our considerations.

By $\mathcal{S}_n$ we denote the symmetric group of order $n$, i.e. the group  of all one-to-one mappings of the  set $[n] =\left\{ 1,2,\ldots ,n\right\}$ in itself. If $x\in [n] $, $\rho\in \mathcal{S}_n$, then the image of the element $x$ in the mapping $\rho$ we denote by $\rho (x)$.

\begin{definition}
Let $g'=\langle R_{g'} ,C_{g'} ,E_{g'} \rangle $ and $g''=\langle R_{g''} ,C_{g''} ,E_{g''} \rangle $ are two bipartite graphs.
We will say that the graphs $g'$ and $g''$ are \emph{isomorphic} and we will write
$$g'\cong g'' ,$$
if $|R_{g'} |=|R_{g''} |=n$, $|C_{g'} |=|C_{g''} |=m$ and there exist $\rho \in {\mathcal S}_{n} $ and $\sigma \in {\mathcal S}_{m} $ such that  $\langle r,c\rangle \in E_{g'} \Leftrightarrow \langle \rho (r),\sigma (c)\rangle \in E_{g''} $.
\end{definition}

In this paper we consider only bipartite graphs up to isomorphism.

For more details on graph theory and its applications see  \cite{diestel,harary}.

The connection between the bipartite graphs and the popular puzzle Sudoku is described in details in  \cite{bigraps-new}. There it is shown that if we want to find the number of all $n^2 \times n^2$ Sudoku grids, it is necessary to obtain all bipartite graphs of the type $g=\langle R_g , C_g , E_g \rangle $ with exactness up to isomorphism and also some of their numerical characteristics, where $|R_g |=|C_g |=n$.

We dedicate this paper on the problem to obtain all bipartite graphs of the type $g=\langle R_g , C_g , E_g \rangle$ with exactness up to isomorphism at given $n$ and $m$, where $n=|R_g |$, $m=|C_g |$. The set of all these graphs we will denote with $\mathfrak{G}_{n\times m}$. For this purpose we will represent  the set $\mathfrak{G}_{n\times m}$ with the help of $n\times m$ binary matrices.

Let us recall that \emph{binary} (or \emph{boolean}, or \emph{(0,1)-matrix}) is called a matrix whose elements belong to the set ${\mathfrak B}=\{ 0,1\} $. With ${\mathfrak B}_{n\times m} $ we will denote the set of all $n\times m$ binary matrices.

A square binary matrix is called a \textit{permutation matrix}, if there is just one 1 in every row and every column. Let us denote by ${\mathcal P}_{n} $ the group of all $n\times n$  permutation matrices. In effect is the isomorphism ${\mathcal P}_{n} \cong {\rm {\mathcal S}}_{n} $.

As it is well known (see \cite{Sachkov,tar2}) that the multiplication of an arbitrary real or complex matrix $A$ from the left with a permutation matrix (if the multiplication is possible) leads to dislocation of the rows of the matrix $A$, while the multiplication of $A$ from the right with a permutation matrix leads to the dislocation of the columns of $A$.

With ${\mathcal T}_{n} \subset {\mathcal P}_{n} $ we denote the set of all \emph{transpositions} in ${\mathcal P}_{n} $, i.e. the set of all $n\times n$ permutation matrices, which multiplying from the left an arbitrary $n\times m$ matrix swaps the places of exactly two rows, while multiplying from the right an arbitrary $k\times n$ matrix swaps the places of exactly two columns.

\begin{definition}\label{Definition3}
Let $A,B\in {\mathfrak B}_{n\times m}$. We will say that the matrices $A$ and $B$ are \emph{equivalent} and we will write
$$A{\rm \sim }B, $$
if there exist permutation matrices $X\in {\mathcal P }_{n} $ and $Y\in {\mathcal P}_{m} $, such that
$$A=XBY.$$

In other words $A \sim B$ if $A$ is received from $B$ after dislocation of some of the rows and the columns of $B$.
Obviously, the introduced relation is an equivalence relation.
\end{definition}

Let $g=\langle R_{g} ,C_{g} ,E_{g} \rangle $ be a bipartite graph, where $R_{g} =\{ r_{1} ,r_{2} ,\ldots ,r_{n} \} $ and $C_{g} =\{ c_{1} ,c_{2} ,\ldots ,c_{m} \} $. Then we build the matrix $A=[a_{ij} ]\in  {\mathfrak B}_{n\times m} $, such that $a_{ij} =1$ if and only if $\langle r_{i} ,c_{j} \rangle \in E_{g} $. Inversely, let $A=[a_{ij} ]\in {\mathfrak B}_{n\times m} $. We denote the $i$-th row of $A$ with $r_{i} $, while the $j$-th column of $A$ with $c_{j} $. Then we build the bipartite graph $g=\langle R_{g} ,C_{g} ,E_{g} \rangle $, where $R_{g} =\{ r_{1} ,r_{2} ,\ldots ,r_{n} \} $, $C_{g} =\{ c_{1} ,c_{2} ,\ldots ,c_{n} \} $ and there exists an edge from the vertex $r_{i} $ to the vertex $c_{j} $ if and only if $a_{ij} =1$. It is easy to see that if $g$ and $h$ are two isomorphic graphs and $A$ and $B$ are the corresponding matrices, then $A$ is obtained from $B$ by a permutation of columns and/or rows.
Thus we showed the following obvious relation between the bipartite graphs and the binary matrices:

\begin{proposition}\label{varphighgh}

There exists one-to-one mapping
$$\varphi : \mathfrak{G}_{n\times m} \to \mathfrak{B}_{n\times m} $$
between the elements of $\mathfrak{G}_{n\times m}$  and  $\mathfrak{B}_{n\times m}$, such that if $g,h\in \mathfrak{G}_n$, then
$$g\cong h \Longleftrightarrow \varphi (g) \sim \varphi (h) .$$

\hfill $\square$
\end{proposition}

Thus, the combinatorial problem to obtain and enumerate all of $n\times m$ binary matrices up to permutation of columns or rows naturally arises.

\section{ Semi-canonical and canonical binary matrices}

Let $A\in {\mathfrak B}_{n\times m} $. With $r(A)$ we will denote the ordered $n$-tuple
$$r(A)=\langle x_{1} ,x_{2} ,\ldots ,x_{n} \rangle ,$$
where $0\le x_{i} \le 2^{m} -1$, $i=1,2,\ldots n$ and $x_{i} $ is a natural number written in binary notation with the help of the $i$-th row of $A$.

Similarly with $c(A)$ we will denote the ordered $m$-tuple
$$c(A)=\langle y_{1} ,y_{2} ,\ldots ,y_{m} \rangle ,$$
where $0\le y_{j} \le 2^{n} -1$, $j=1,2,\ldots m$ and $y_{j} $ is a natural number written in binary notation with the help of the $j$-th column of $A$.

We consider the sets:

\[\begin{array}{lll} {{\mathcal R}_{n\times m} } & {=} & {\left\{\langle x_{1} ,x_{2} ,\ldots ,x_{n} \rangle \; |\; 0\le x_{i} \le 2^{m} -1,\; i=1,2,\ldots n\right\}} \\ {} & {=} & {\left\{r(A)\, |\; A\in {\mathfrak B}_{n\times m} \right\}} \end{array}\]
and

\[\begin{array}{lll} {{\mathcal C}_{n\times m} } & {=} & {\left\{\langle y_{1} ,y_{2} ,\ldots ,y_{m} \rangle \; |\; 0\le y_{j} \le 2^{n} -1,\; j=1,2,\ldots m\right\}} \\ {} & {=} & {\left\{c(A)\, |\; A\in {\mathfrak B}_{n\times m} \right\}} \end{array}\]

Thus we define the following two mappings:
$$r: {\mathfrak B}_{n\times m} \to {\mathcal R}_{n\times m} $$
and
$$c: {\mathfrak B}_{n\times m} \to  {\mathcal C}_{n\times m} ,$$
which are bijective and therefore
$${\mathcal R}_{n\times m} \cong {\mathfrak B}_{n\times m} \cong {\rm {\mathcal C}}_{n\times m} .$$

The above described bijections $r$ and $c$  leads to the following statement, which is an analog of Proposition \ref{varphighgh}:

\begin{proposition}\label{prop2ghgh}
There exist one-to-one mappings between the elements of $\mathfrak{G}_{n\times m}$  and  the sets
$${\mathcal R}_{n\times m} = \left\{\langle x_{1} ,x_{2} ,\ldots ,x_{n} \rangle \; |\; 0\le x_{i} \le 2^{m} -1,\; i=1,2,\ldots n\right\}$$
and

$${\mathcal C}_{n\times m} = \left\{\langle y_{1} ,y_{2} ,\ldots ,y_{m} \rangle \; |\; 0\le y_{j} \le 2^{n} -1,\; j=1,2,\ldots m\right\}$$

\hfill $\square$
\end{proposition}

\begin{example}\label{primergraph}
The shown in Figure \ref{n3k9} graph is unambiguously coded with the help of the matrix
$$A=\left[\begin{array}{cccc} {1} & {1} & {0} & {0} \\ {1} & {1} & {1} & {0} \\ {0} & {0} & {0} & {1}  \end{array}\right]$$
and the help of the ordered set of nonnegative integers.
$$r(A)=\langle 12, 14, 1 \rangle$$
and
$$c(A) = \langle 6, 6, 2, 1 \rangle .$$

\unitlength=0.8mm \linethickness{0.6pt}

\begin{figure}
\begin{center}
\begin{picture}(50,50)

\put(0,0){\framebox(50,50)}
\put(10,43){\makebox(0,0){$g$}}

\put(13,21){\oval(14,30)}
\put(37,26){\oval(14,40)}

\put(14,3){\makebox(0,0){$R_g$}}
\put(38,3){\makebox(0,0){$C_g$}}

\put(37,40){\circle{2}}

\put(13,30){\circle*{2}}
\put(37,30){\circle{2}}
\put(15,30){\line(2,1){20}}
\put(15,30){\line(1,0){20}}

\put(13,20){\circle*{2}}
\put(37,20){\circle{2}}
\put(15,20){\line(1,1){20}}
\put(15,20){\line(1,0){20}}
\put(15,20){\line(2,1){20}}

\put(13,10){\circle*{2}}
\put(37,10){\circle{2}}
\put(15,10){\line(1,0){20}}
\end{picture}
\end{center}
\caption{}\label{n3k9}
\end{figure}

\end{example}

The lexicographic orders in ${\rm {\mathcal R}}_{n\times m} $ and in ${\rm {\mathcal C}}_{n\times m}$  we will denote with $<$.

\begin{definition}\label{Definition2}
\rm Let $A\in {\mathfrak B}_{n\times m} $,
$r(A)=\langle x_{1} ,x_{2} ,\ldots ,x_{n} \rangle$ and
$c(A)=\langle y_{1} ,y_{2} ,\ldots ,y_{m} \rangle$.
We will call the matrix $A$ \emph{semi-canonical}, if
$$x_{1} \le x_{2} \le \cdots \le x_{n} $$
and
$$y_{1} \le y_{2} \le \cdots \le y_{m} .$$
\end{definition}

\begin{proposition}\label{Proposition1}
Let $A=[a_{ij} ]\in {\mathfrak B}_{n\times m} $ be a semi-canonical matrix. Then there exist integers $i,j$, such that $1\le i\le n$, $1\le j\le m$ and
\begin{equation} \label{GrindEQ__2_}
 a_{1\; {\kern 1pt} 1} =a_{1{\kern 1pt} \; 2} =\cdots =a_{1{\kern 1pt} \; j} =0,\quad a_{1{\kern 1pt} \; j+1} =a_{1{\kern 1pt} \; j+2} =\cdots =a_{1{\kern 1pt} \; m} =1, \end{equation}
\begin{equation} \label{GrindEQ__3_}
a_{1\; {\kern 1pt} 1} =a_{2{\kern 1pt} \; 1} =\cdots =a_{i{\kern 1pt} \, \; 1} =0,\quad a_{i+1\, {\kern 1pt} \; 1} =a_{i+2{\kern 1pt} \; \, 1} =\cdots =a_{n\; \, {\kern 1pt} 1} =1. \end{equation}
\end{proposition}

Proof. Let $r(A)=\langle x_{1} ,x_{2} ,\ldots x_{n} \rangle $ and $c(A)=\langle y_{1} ,y_{2} ,\ldots y_{m} \rangle $. We assume that there exist integers $p$ and $q$, such that $1\le p<q\le m$, $a_{1{\kern 1pt} p} =1$ and $a_{1{\kern 1pt} q} =0$. In this case $y_{p} >y_{q} $, which contradicts the condition for semi-canonicity of the matrix $A$. We have proven (\ref{GrindEQ__2_}). Similarly, we prove (\ref{GrindEQ__3_}) as well.

\hfill $\square$

\begin{corollary} \label{Corollary1}
Let $A=[a_{ij} ]\in {\mathfrak B}_{n\times m} $ be a semi-canonical matrix. Then there exist integers $s,t$, such that $0\le s\le m$, $0\le t\le n$, $x_{1} =2^{s} -1$ and $y_{1} =2^{t} -1$

\hfill $\square$
\end{corollary}

\begin{definition}\label{Definition4}
\rm We will call the matrix $A\in {\mathfrak B}_{n\times m} $ \emph{canonical matrix}, if $r(A)$ is a minimal element about the lexicographic order in the set $\overline{A}= \{ r(B)\; |\; B \sim A\} $.
\end{definition}

If the matrix $A\in {\mathfrak B}_{n\times m} $ is canonical and $r(A)=\langle x_{1} ,x_{2} ,\ldots ,x_{n} \rangle ,$ then obviously

\begin{equation} \label{GrindEQ__6_} x_{1} \le x_{2} \le \cdots \le x_{n} . \end{equation}

From Definition \ref{Definition4} immediately follows that in every equivalence class about the relation $"\sim "$ (see Definition \ref{Definition3}) there exists only one canonical matrix. Therefore, to find all bipartite graphs of type $g=\langle R_{g} ,C_{g} ,E_{g} \rangle $, $|R_{g} |=n$, $|C_{g} |=m$ up to isomorphism, it suffices to find all canonical matrices from the set ${\mathfrak B}_{n\times m} $.

\begin{theorem}\label{Theorem1}
Let $A$ be an arbitrary matrix from ${\mathfrak B}_{n\times m} $. Then:

a) If $X_{1} ,X_{2} ,\cdots ,X_{s} \in {\rm {\mathcal T}}_{n} $ are such that
$$r(X_{1} X_{2} \ldots X_{s} A)<r(X_{2} X_{3} \ldots X_{s} A)<\cdots <r(X_{s} A)<r(A),$$
then
$$c(X_{1} X_{2} \ldots X_{s} A)<c(A).$$

b) If $Y_{1} ,Y_{2} ,\cdots ,Y_{t} \in {\rm {\mathcal T}}_{m} $ are such that
$$c(AY_{1} Y_{2} \ldots Y_{t})<c(AY_{2} Y_{3} \ldots Y_{t})<\cdots <c(AX_{t})<r(A),$$
then
$$r(AY_{1} Y_{2} \ldots Y_{t} )<r(A).$$
\end{theorem}

Proof. a) Induction by $s$.

Let $s=1$ and let $X\in {\mathcal T}_{n} $ be a transposition which multiplying an arbitrary matrix $A=[a_{ij} ]\in {\mathfrak B}_{n\times m} $ from the left swaps the places of the rows of $A$ with numbers $u$ and $v$ ($1\le u<v\le n$), while the remaining rows stay in their places. In other words if
$$A=\left[
  \begin{array}{cccccc}
    a_{11} & a_{12} & \cdots & a_{1r} & \cdots & a_{1m} \\
    a_{21} & a_{22} & \cdots & a_{2r} & \cdots & a_{2m} \\
    \vdots & \vdots &        & \vdots &        & \vdots \\
    a_{u1} & a_{u2} & \cdots & a_{ur} & \cdots & a_{um} \\
    \vdots & \vdots &        & \vdots &        & \vdots \\
    a_{v1} & a_{v2} & \cdots & a_{vr} & \cdots & a_{vm} \\
    \vdots & \vdots &        & \vdots &        & \vdots \\
    a_{n1} & a_{n2} & \cdots & a_{nr} & \cdots & a_{nm} \\
  \end{array}
\right]
$$
then
$$X_1 A=\left[
  \begin{array}{cccccc}
    a_{11} & a_{12} & \cdots & a_{1r} & \cdots & a_{1m} \\
    a_{21} & a_{22} & \cdots & a_{2r} & \cdots & a_{2m} \\
    \vdots & \vdots &        & \vdots &        & \vdots \\
    a_{v1} & a_{v2} & \cdots & a_{vr} & \cdots & a_{vm} \\
    \vdots & \vdots &        & \vdots &        & \vdots \\
    a_{u1} & a_{u2} & \cdots & a_{ur} & \cdots & a_{um} \\
    \vdots & \vdots &        & \vdots &        & \vdots \\
    a_{n1} & a_{n2} & \cdots & a_{nr} & \cdots & a_{nm} \\
  \end{array}
\right] ,
$$
where $a_{ij} \in\{ 0,1\}$, $1\le i\le n$, $1\le j\le m$.

Let $$r(A)=\langle x_{1} ,x_{2} ,\ldots ,x_{u} ,\ldots ,x_{v} ,\ldots ,x_{n} \rangle .$$

Then $$r(XA)=\langle x_{1} ,x_{2} ,\ldots ,x_{v} ,\ldots ,x_{u} ,\ldots ,x_{n} \rangle .$$

Since $r(XA)<r(A)$, then according to the properties of the lexicographic order $x_{v} <x_{u} $. Let the representation of $x_{u} $ and $x_{v} $ in binary notation with an eventual addition if necessary with unessential zeros in the beginning be respectively as follows:
$$x_{u} =a_{u1} a_{u2} \cdots a_{um} ,$$
$$x_{v} =a_{v1} a_{v2} \cdots a_{vm} .$$

Since $x_{v} <x_{u} $, then there exists an integer $r\in \{ 1,2,\ldots ,m\} $, such that $a_{uj} =a_{vj} $ when $j<r$, $a_{ur} =1$ and $a_{vr} =0$.
Hence if $c(A)=\langle y_{1} ,y_{2} ,\ldots ,y_{m} \rangle $, $c(XA)=\langle z_{1} ,z_{2} ,\ldots ,z_{m} \rangle $, then $y_{j} =z_{j} $ when $j<r$, while the representation of $y_{r} $ and $z_{r} $ in binary notation with an eventual addition if necessary with unessential zeroes in the beginning is respectively as follows:
$$y_{r} =a_{1r} a_{2r} \cdots a_{u-1{\kern 1pt} r} a_{ur} \cdots a_{vr} \cdots a_{nr} ,$$
$$z_{r} =a_{1r} a_{2r} \cdots a_{u-1{\kern 1pt} r} a_{vr} \cdots a_{ur} \cdots a_{nr} .$$

Since $a_{ur} =1$, $a_{vr} =0$, then $z_{r} <y_{r} $, whence it follows that $c(XA)<c(A)$.

We assume that for every $s$-tuple of transpositions $X_{1} ,X_{2} ,\ldots ,X_{s} \in {\mathcal T}_{n} $ and for every matrix $A\in {\mathfrak B}_{n\times m} $ from
$$r(X_{1} X_{2} \ldots X_{s} A)<r(X_{2} \cdots X_{s} A)<\cdots <r(X_{s} A)<r(A)$$
it follows that
$$c(X_{1} X_{2} \ldots X_{s} A)<c(A)$$
and let $X_{s+1} \in {\mathcal T}_{n} $ be such that
$$r(X_{1} X_{2} \ldots X_{s} X_{s+1} A)<r(X_{2} \cdots X_{s+1} A)<\cdots <r(X_{s+1} A)<r(A).$$

According to the induction assumption $c(X_{s+1} A)<c(A)$.

We put

$$A_{1} =X_{s+1} A.$$

According to the induction assumption from
$$r(X_{1} X_{2} \ldots X_{s} A_{1} )<r(X_{2} \cdots X_{s} A_{1} )<\cdots <r(X_{s} A_{1} )<r(A_{1} )$$
it follows that
$$c(X_{1} X_{2} \cdots X_{s} X_{s+1} A)=c(X_{1} X_{2} \cdots X_{s} A_{1} )<c(A_{1} )=c(X_{s+1} A)<c(A),$$
with which we have proven a).

b) is proven similarly to a).

\hfill $\square$

Obviously in effect is also the dual to Theorem \ref{Theorem1} statement, in which everywhere instead of the sign $<$ we put the sign $>$.

\begin{corollary}\label{Corollary2}
If the matrix $A\in {\mathfrak B}_{n\times m} $  is a canonical matrix, then $A$ is a semi-canonical matrix.
\end{corollary}

Proof. Let $A\in {\mathfrak B}_{n\times m} $ be a canonical matrix and $r(A)=\langle x_{1} ,x_{2} ,\ldots ,x_{n} \rangle $. Then from (\ref{GrindEQ__6_}) it follows that $x_{1} \le x_{2} \le \cdots \le x_{n} $. Let $c(A)=\langle y_{1} ,y_{2} ,\ldots ,y_{m} \rangle $. We assume that there are $s$ and $t$ such that $s\le t$ and $y_{s} >y_{t} $. Then we swap the columns of numbers $s$ and $t$. Thus we obtain the matrix $A'\in {\mathfrak B}_{n\times m} $, $A'\ne A$. Obviously $c(A')<c(A)$. From Theorem \ref{Theorem1} it follows that $r(A')<r(A),$ which contradicts the minimality of $r(A)$.

\hfill $\square$

In the next example, we will see that the opposite statement of Corollary 2 is not always true.

\begin{example}\label{Example1}
\rm We consider the matrices:
$$A=\left[\begin{array}{cccc} {0} & {0} & {1} & {1} \\ {0} & {0} & {1} & {1} \\ {0} & {1} & {0} & {0} \\ {1} & {0} & {0} & {0} \end{array}\right]$$
and
$$B=\left[\begin{array}{cccc} {0} & {0} & {0} & {1} \\ {0} & {1} & {1} & {0} \\ {0} & {1} & {1} & {0} \\ {1} & {0} & {0} & {0} \end{array}\right] .$$

After immediate verification, we find that $A\sim B$. Furthermore $r(A)=\langle 3,3,4,8\rangle $, $c(A)=\langle 1,2,12,12\rangle $, $r(B)=\langle 1,6,6,8\rangle $, $c(B)=\langle 1,6,6,8\rangle $. So $A$ and $B$ are two equivalent to each other semi-canonical matrices, but they are not canonical. Canonical matrix in this equivalence class is the matrix
$$C=\left[\begin{array}{cccc} {0} & {0} & {0} & {1} \\ {0} & {0} & {1} & {0} \\ {1} & {1} & {0} & {0} \\ {1} & {1} & {0} & {0} \end{array}\right],$$
where $$r(C)=\langle 1,2,12,12\rangle ,\quad c(C)=\langle 3,3,4,8\rangle .$$

\hfill $\square$
\end{example}

From example \ref{Example1} immediately follows that in a given equivalence class it is possible to exist more than one semi-canonical element.

In \cite{fmns2015} we described and we implemented with help of C++ programming language an algorithm for finding all $n\times n$ semi-canonical binary matrices taking into account the number of 1 in each of them. In the described algorithm, the bitwise operations are substantially used.

Let us denote with $\beta (n,k)$ the number of all $n\times n$ semi-canonical binary matrices with exactly $k$ 1's, where $0\le k\le n^{2} $. In \cite{fmns2015}, we received the following integer sequences:
$$ \left\{\beta (2,k)\right\}_{k=0}^{4} =\left\{1, 1, 3, 1, 1\right\}$$
$$ \left\{\beta (3,k)\right\}_{k=0}^{9} =\left\{1, 1, 3, 8, 10, 9, 8, 3, 1, 1\right\}$$
$$ \left\{\beta (4,k)\right\}_{k=0}^{16} =\left\{1, 1, 3, 8, 25, 49, 84, 107, 121, 101, 72, 41, 24, 8, 3, 1, 1\right\}$$
$$\left\{\beta (5,k)\right\}_{k=0}^{25} =\{ 1, 1, 3, 8, 25, 80, 220, 524, 1057, 1806, 2671, 3365, 3680, 3468,$$
$$2865, 2072, 1314, 723, 362, 166, 72, 24, 8, 3, 1, 1\} $$
$$\left\{\beta (6,k)\right\}_{k=0}^{36} =\{ 1, 1, 3, 8, 25, 80, 283, 925, 2839, 7721, 18590, 39522, 74677, $$
$$125449, 188290, 252954, 305561, 332402, 326650, 290171, 233656, 170704, $$
$$113448, 68677, 37996, 19188, 8910, 3847, 1588, 613, 299, 72, 24, 8, 3, 1, 1\}$$

\section{A necessary and sufficient condition for a binary matrix to be canonical}

Let $A=[a_{ij} ]\in \mathfrak{B}_{n\times m}$, $r(A)=\langle x_1 ,x_2 , \ldots , x_n \rangle$. We denote the following notations:

\begin{description}
\item[$\varepsilon_i (A)$]$\displaystyle=\varepsilon (x_i ) =\sum_{j=1}^m a_{ij}$ -- the number of 1 in the $i$-th row of $A$, $i=1,2,\ldots n$.

\item[$Z_i (A)$]$\displaystyle =Z(x_i ) =\{ x_k \in r(A)\; |\; x_k =x_i \}$ -- the set of all rows, equal to $i$-th row of $A$. By definition $x_i \in Z (x_i )$, $i=1,2,\ldots n$.

\item[$\zeta_i (A)$]$\displaystyle=\zeta (x_i ) = |Z_i (A) |$, $i=1,2,\ldots n$.
\end{description}

The next four statements are obvious and their proof is trivial.
\begin{proposition}\label{tv4}
Let $A=[a_{ij} ]\in \mathfrak{B}_{n\times m}$, $r(A)=\langle x_1 ,x_2 , \ldots , x_n \rangle$  and let $x_1 \le x_2 \le \cdots \le x_n$. Let $s$ и $t$ are integers so that $0\le s< n$, $0\le t< m$ and let the matrix $B\in \mathfrak{B}_{(n-t)\times (m-s)}$ be obtained from $A$ removing the first $t$ rows and the last $s$ columns and let $r(B)=\langle x_{t+1}',x_{t+2}',\ldots ,x_{n}' \rangle$. Then $x_{t+1}'\le x_{t+2}'\le \cdots \le x_{n}'$.

\hfill $\square$
\end{proposition}

\begin{proposition}\label{tv5}
Let $A=[a_{ij} ]\in \mathfrak{B}_{n\times m}$, $r(A)=\langle x_1 ,x_2 , \ldots , x_n \rangle$ and let $x_1 \le x_2 \le \cdots \le x_n$. Then for each $i=2,3,\ldots ,n$, for which $x_{i-1} <x_i$, or $i=1$ the condition
$$ Z(x_i )=\{ x_i ,x_{i+1} ,\ldots , x_{i+\zeta (x_i ) } \} $$
is fulfilled.

\hfill $\square$
\end{proposition}

\begin{proposition}\label{tv6}
Let $A=[a_{ij} ]\in \mathfrak{B}_{n\times m}$, $r(A)=\langle x_1 ,x_2 , \ldots , x_n \rangle$,  $x_1 \le x_2 \le \cdots \le x_n$ and let $\varepsilon_1 (A) =m$, i.e. $x_1 =2^m-1$. Then $A$ is canonical.

\hfill $\square$
\end{proposition}

\begin{proposition}\label{tv7}
 Let $A=[a_{ij} ]\in \mathfrak{B}_{n\times m}$, $r(A)=\langle x_1 ,x_2 , \ldots , x_n \rangle$, $x_1 =2^s -1$  for some integer $s$, $0\le s\le m$ and let $\zeta_1 (A) =n$. Then $A$ is canonical.

\hfill $\square$
\end{proposition}

\begin{theorem}\label{mainth}
Let $A=[a_{ij} ]\in \mathfrak{B}_{n\times m}$, $$r(A)=\langle x_1 ,x_2 , \ldots , x_n \rangle ,$$ $$c(A)=\langle y_1 ,y_2 ,\ldots ,y_m \rangle.$$ Then $A$ is canonical if and only if  the next condition are true:
\begin{enumerate}
  \item \label{cond1} $x_1 \le x_2 \le \cdots \le x_n$;
  \item \label{cond2} $x_1 =2^s -1$, where $s=\varepsilon_1(A)$;
  \item \label{cond3} For each  $i=2,3,\ldots ,n$ is fulfilled $\varepsilon_1 (A)\le \varepsilon_i (A)$;
  \item \label{cond4} If for some integer $i$ such that $\zeta_1 (A) <i\le n$ is fulfilled $\varepsilon_i (A)=\varepsilon_1 (A)$, then $\zeta_1 (A) \ge \zeta_i (A)$;
  \item \label{cond5} $y_{m-s+1} \le y_{m-s+2} \le \cdots \le y_m$, where $s=\varepsilon_1 (A)$;
  \item \label{cond6}  Let for some integer $i$ such that $\zeta_1 (A) <i\le n$ are fulfilled  $\varepsilon_i (A)=\varepsilon_1 (A)=s$ and $\zeta_1 (A) =\zeta_i (A)=t$. Let $\Upsilon (x_i )=\{ y_i \in c(A)\; |\; a_{ij} =1\} =\{y_{u_1} , y_{u_2 } ,\ldots ,y_{u_s} \}$. Let $A'$ is the matrix which is obtained from $A$ replacing the places of the rows from the set $Z(x_1 )$ with the rows of the set $Z(x_i )$, the place of the column  $y_{u_1}$ with the place of the column $y_{m-s}$, the place of the column $y_{u_2}$ with the place of the column $y_{m-s+1}$ and so on, the place of the column  $y_{u_s}$  with the place of the column $y_{m}$.\footnote{If for some $j$ is satisfied $u_j = m-s+j$, then $y_{u_j}$ remains at its place.} Then $r(A)\le r(A')$;
  \item \label{cond7} If $s=\varepsilon_1 (A)<m$ and  $t=\zeta_1 (A)<n$ then the matrix $B\in \mathfrak{B}_{(n-s)\times (m-t)}$, which is obtained from $A$ removing the first $s$ rows and the last $t$ columns is canonical.
\end{enumerate}
\end{theorem}

Proof.

Necessity. Let $A=[a_{ij} ]\in \mathfrak{B}_{n\times m}$ be a canonical matrix and let $r(A)=\langle x_1 ,x_2 , \ldots , x_n \rangle$, $c(A)=\langle y_1 ,y_2 ,\ldots ,y_m \rangle$.

Conditions \ref{cond1} and \ref{cond5} are due to the fact that every canonical matrix is semi-canonical (Corollary \ref{Corollary2}), so $x_1 \le x_2 \le \cdots \le x_n$ и $y_1 \le y_2 \le \cdots \le y_m$.

Condition \ref{cond2} comes from Corollary \ref{Corollary1}.

\ref{cond3}. We assume that an integer $i$, $2\le i\le n$ exists, such that $\varepsilon_i (A)<\varepsilon_1 (A)=s$ and let $\varepsilon_i (A)=u<s$. Then a matrix $A'=[a_{i\, j}']\sim A$ exists such that $a_{i\, 1}'=a_{i\> 2}' =\cdots =a_{i\, m-u}' =0 $ and $a_{i\, m-u+1}' = a_{i\, m-u+2}' =\cdots = a_{i\, m} =1$. We move the $i$-th row  of $A'$ at first place and we obtain a matrix $A'' $. Obviously $A'' \sim A$. Let $r(A'' )=\langle x_1'' ,x_2'' ,\ldots ,x_n'' \rangle$. Then $x_1'' =2^u -1<2^s -1=x_1$. Therefore $r(A'' )<r(A)$, which is impossible, due to the fact that $A$ is canonical.

\ref{cond4}. Let $s=\varepsilon_1 (A)$ and $t=\zeta_1 (A)$. According to the proved above  condition  \ref{cond2} and Proposition \ref{tv5} we have $x_1 = x_2 = \cdots = x_t = 2^s -1 <x_{t+1}$. We assume that an integer $i$, $t<i\le n$ exists, such that $\varepsilon_i (A)=\varepsilon_1 (A)=s$ and $\zeta_i (A)>\zeta_1 (A)$. Let $\zeta_i (A)=v$, $v>t$. Then a matrix $A' \sim A$ exist, such that $r(A')=\langle x_1',x_2',\ldots , x_n' \rangle$, where $x_1'=x_2'=\cdots =x_v' =2^s -1<x_{v+1}$. Due to the fact that $t+1\le v$, consequently $x_{t+1}' =x_v' =2^s -1 =x_t <x_{t+1}$. We obtained that $x_k =x_k'$ for $k=1,2,\ldots ,t$ and  $x_{t+1}' < x_{t+1}$. From here it follows that $r(A')<r(A)$, which is contrary to the canonicity of $A$.

Condition \ref{cond6} comes directly from the fact that $A$ is canonical and $r(A)\le r(A')$ for each matrix $A'\sim A$.

\ref{cond7}. From the already proved condition \ref{cond1} $\div$ \ref{cond4} and Proposition \ref{tv5} it follows that $A$ is presented in this type:
\begin{equation}\label{kiki}
A= \left[
  \begin{array}{cc}
    O & E \\
    B & C \\
  \end{array}
\right] ,
\end{equation}
where $O$ is $t\times (m-s)$ matrix, all element of which are equal to 0, $E$ is $t\times s$ matrix, all element of which are equal to 1, $B \in \mathfrak{B}_{(n-t)\times (m-s)}$, as the first rows of $B$ are not entirely null, $C \in \mathfrak{B}_{(n-t)\times s}$, $s=\varepsilon_1 (A)$ and $t=\zeta_1 (A)$.

Let $B' \sim B$ and $B'$ is $(n-t)\times (m-s)$ canonical binary matrix. Then the following matrices $A' \in \mathfrak{B}_{n\times n}$ and $C'\in \mathfrak{B}_{(n-t)\times s}$ exist, such that $A' \sim A$, $C' \sim C$, $\displaystyle A'= \left[
  \begin{array}{cc}
    O & E \\
    B' & C' \\
  \end{array}
\right] $, and $C'$ is obtained from $C$ after eventual change some of the rows.
Let $r(B) =\langle b_1 ,b_2 ,\ldots , b_{m-s} \rangle$, $r(B') =\langle b_1' ,b_2' ,\ldots , b_{m-s}' \rangle$, $r(C) =\langle c_{m-s+1} ,c_{m-s+2} ,\ldots , c_{m} \rangle$, $r(C') =\langle c_{m-s+1}' ,c_{m-s+2}' ,\ldots , c_{m}' \rangle$, $r(A')=\langle x_1' ,x_2' ,\ldots , x_n' \rangle$. Obviously $x_i'= x_i =2^s -1$ for each $i=1,2,\ldots ,t$. Because $B'$ is canonical, and $0\le c_k , c_k' <2^s$ for each $k\in [m-s+1, m]$ there exist $i\in \{t+1 , t+2 ,\ldots ,n\}$ such that $b_1' =b_1, b_2' =b_2 ,\ldots , b_{i-1}' =b_{i-1}$ and  $b_i' < b_i$. Then $x_1' =x_1 ,x_2' =x_2 ,\ldots ,x_{i-1}' =x_{i-1}$ and $x_i' =b_i' 2^s +c_i' \le b_i 2^s +c_i =x_i$. Consequently $r(A')\le r(A)$. But $A$ is canonical, i.e.  $r(A)\le r(A')$. Therefore  $A'=A$, from where $B'=B$  and $B$ is canonical.

Sufficiency. Let $A\in\mathfrak{B}_{n\times m}$ satisfy the conditions \ref{cond1} $\div$ \ref{cond7} and let  $A'\in\mathfrak{B}_{n\times m}$ be a canonical matrix, $A' \sim A$. Since the conditions 1 $\div$ 7 are necessary for the canonicity of a matrix, consequently $A'$ also satisfies these conditions.

For $A' \sim A$ and having in mind conditions \ref{cond2} $\div$ \ref{cond4} it is easy to see that
\begin{equation}\label{star1}
\varepsilon_1 (A')=\varepsilon_1 (A)=s \quad \textrm{and} \quad \zeta_1 (A') =\zeta_1 (A)=t .
\end{equation}

If $s=m$, according to Proposition \ref{tv6} the matrix $A$ is canonical. If $t=n$, according to Proposition \ref{tv7} the matrix $A$ is canonical.

Let $1\le s< m$ и $1\le t<n$. In this case conditions \ref{cond1} $\div$ \ref{cond4}, Proposition \ref{tv5} and equations (\ref{star1}) guarantee that $A$ and $A'$ are presented in the type
\begin{equation}\label{star2}
A= \left[
  \begin{array}{cc}
    O & E \\
    B & C \\
  \end{array}
\right]
\quad \textrm{and} \quad
A'= \left[
  \begin{array}{cc}
    O & E \\
    B' & C' \\
  \end{array}
\right] ,
\end{equation}
where $O$ is $t\times (m-s)$ matrix, all elements of which are equal to 0, $E$ is $t\times s$ matrix, all elements of which are equal to 1, $B,B' \in \mathfrak{B}_{(n-t)\times (m-s)}$ as the first rows of $B$ and $B'$ are not entirely null and $C,C' \in \mathfrak{B}_{(n-t)\times s}$.

According to conditions \ref{cond5} and \ref{cond7} and the equations (\ref{star2}), it is easy to see that if $c(A) =\langle y_1 ,y_2 ,\ldots ,y_m \rangle$ and $c(A') =\langle y_1' ,y_2' ,\ldots ,y_m' \rangle$, then $y_1 \le y_2 \le\cdots \le y_m $ and $y_1' \le y_2' \le \cdots \le y_m' $. Therefor the matrices $A$ and $A'$ are semi-canonical.

We assume that $A'\ne A $ and $A'$ we obtain that the change of $A$ of some of the columns. Let as change the places of column with numbers $k$ and $l$, $1\le k < l\le m$. The inequality $m-s< k< l\le m$, where $s=\varepsilon_1 (A)=\varepsilon_1 (A')$ is impossible due the condition \ref{cond5}. The inequality $1\le k<l\le m-s$  is impossible due the condition \ref{cond7}. Consequently $1\le k\le m-s<l\le m$. But then having in mind (\ref{star2}), it is easy to see that that in this case it is necessary also to change the places of some of the rows of $A$.

Let $A'$ be obtained after changing the places of some of the rows and afterwards possibly of some of of the columns of $A$. Let us change the places of the rows with the number $i$ and $j$ of the matrix  $A$, where $1\le i<j\le n$. If  $1\le i<j\le t=\zeta_1 (A) =\zeta_1 (A')$, the change of these rows does not lead to alteration of the matrix. If $t<i<j\le n$, then the condition \ref{cond7} will be broken. So $1\le i\le t< j\le n$. According to conditions \ref{cond2} $\div$ \ref{cond4}, $\varepsilon_j (A)=\varepsilon_j (A')=s$ and $\zeta_j (A) = \zeta_j (A')=t$. Consequently  we have changed the place of the first equal to each other $t=\zeta_1 (A)$ rows with another equal to each rows of the set $Z_j (A)$. After that in order to obtain a matrix of kind (\ref{kiki}) it is necessary to change the places of some columns of the matrix $A$. But this contrary to condition \ref{cond6} and to the assumption that $A'\ne A$ and $A'$ is canonical.

Therefore $A=A'$, i.e. $A$ is canonical.

\hfill $\square$

\section{Conclusions and future work} The formulation of Theorem \ref{mainth} is a good basis for the creation of an algorithm receiving all  $n\times m$ canonical binary matrices, which on the other hand describe all bipartite graphs (Proposition \ref{varphighgh}) of the type  $g=\langle R_g , C_g , E_g \rangle $ up to isomorphism, where $|R_g |=n$, $C_g = m$. We will have to settle this problem in the near futures. This paper will be very useful for its solving.


\end{document}